\documentclass{PoS}
\title{Pion mass difference from vacuum polarization}
\ShortTitle{Pion mass difference from vacuum polarization}
\author{\speaker{E. Shintani}$^1$\thanks{shintani@post.kek.jp}, H. Fukaya$^2$, S. Hashimoto$^{1,3}$, 
        H. Matsufuru$^1$, J. Noaki$^1$, T. Onogi$^4$, N. Yamada$^{1,3}$ (for JLQCD Collaboration)\\
        $^1$High Energy Accelerator Research Organization (KEK), Tsukuba 305-0801, Japan\\
        $^2$Theoretical Physics Laboratory, RIKEN, Wako 351-0198, Japan,\\
        $^3$School of High Energy Accelerator Science, 
            The Graduate University for Advanced Studies (Sokendai), Tsukuba 305-0801, Japan,\\
        $^4$Yukawa Institute for Theoretical Physics, Kyoto University, Kyoto 606-8502, Japan}

\abstract{
We calculate the electromagnetic contribution to the pion mass difference, 
$\Delta m^2_\pi=m^2_{\pi^+}-m^2_{\pi^0}$, in the chiral limit through the
$VV-AA$ type vacuum polarization using
Das-Guralnik-Mathur-Low-Young (DGMLY) sum rule.
The calculation is made with two-flavors of dynamical overlap fermions on a
$16^3\times 32$ lattice at $a\sim$0.12 fm.
The exact chiral symmetry of the overlap fermion is essential to control 
the systematic error in the difference $VV-AA$. 
We obtain $\Delta m_\pi^2 = 1024(100)\,{\rm MeV^2}$ combining the lattice data with the perturbative 
contribution in the high momentum region evaluated by the operator product expansion.
By analyzing the momentum dependence of the vacuum polarization, we also obtain 
pion decay constant $f_\pi$ and the low-energy constants $L_{10}^r$ in the chiral limit.}

\FullConference{The XXV International Symposium on Lattice Field Theory\\
         July 30 - August 4 2007\\
         Regensburg, Germany}

\begin{document}

\section{Introduction}
The mass difference between charged pion ($\pi^+$) and neutral pion
($\pi^0$) is considered to be dominated by the electromagnetic (EM)
contribution, 
which contains non-perturbative physics through the (off-shell) pion-pion-photon vertex.
Determination of this quantity with good precision is necessary to determine 
up and down quark masses, which can be done in principle using lattice QCD.

In 1967, Das, Guralnik, Mathur, Low and Young \cite{DGMLY} derived a sum rule, 
which relates the difference between the spectral functions for vector and axial-vector
currents to $\Delta m_\pi^2=m_{\pi^+}^2-m_{\pi^0}^2$ (DGMLY sum rule), which is a generalization 
of the Weinberg sum rule \cite{Weinberg}.
By assuming a saturation by the lowest resonance states of 
vector (rho meson) and axial-vector ($a_1$ meson), their estimate of $\Delta m_\pi^2$ was
already close to the experimental value, $\Delta m_\pi^2({\rm Exp.})=1261.2$ MeV$^2$ \cite{PDG},
which implies that the non-perturbative contributions are important.
There have also been results from extended chiral perturbation
theory including resonance states \cite{Ecker} or Bethe-Salpeter equation \cite{Harada}.
In lattice QCD, $\Delta m_\pi^2$ has been calculated in quenched QCD
with the Wilson fermion \cite{Duncan} and $N_f=2$ QCD with domain-wall fermion \cite{Blum}.
In these works, the EM interaction is introduced to make the
(QCD+QED) system on the lattice, and the charged and neutral
pseudoscalar meson masses are calculated through two-point functions in the usual way.
Their values also show good agreement with $\Delta m_\pi^2({\rm Exp.})$ within the error.

In this work, we apply the DGMLY sum rule to the evaluation of $\Delta m_\pi^2$
(for an early attempt, see Ref. \cite{Gupta}).
With the DGMLY sum rule, $\Delta m_\pi^2$ is written in terms of a
momentum integral of the difference of the vacuum polarizations defined
by the vector and axial-vector currents. 
There are two points to be noted.
First, the DGMLY sum rule exactly holds only in the chiral limit, and hence
the lattice calculation requires good control of the chiral expansion. 
Second, the vector and axial-vector currents form a chiral multiplet 
in the continuum theory.
Their difference signals the spontaneous breaking of chiral symmetry.
With the domain-wall fermions the chiral symmetry is not good enough to
calculate $\langle VV - AA \rangle$ unless the depth in the fifth dimension is
unusually large.
By using the overlap fermion, the chiral symmetry exactly holds, 
which makes the extraction of $\Delta m_\pi^2$ possible. 

The $VV-AA$ vacuum polarization also provides pion decay constant $f_\pi$ 
and a low-energy constants (LECs) $L_{10}^r$.
We also present a calculation of these quantities with two-flavor dynamical
overlap fermion. 

\section{Definition}
\subsection{Continuum formula}
The leading order EM contribution to the pion mass difference is given
by one photon exchange diagram in the self-energy calculation of the
charged and neutral pions as
\begin{eqnarray}
\Delta m_\pi^2 = \int\frac{d^4 q}{(2\pi)^4}\frac{1}{2} D_{\mu\nu}(q)\int d^4 x e^{iqx}
\Big[ \langle \pi^+| T\{J^{\rm EM}_\mu,J^{\rm EM}_\nu\}|\pi^+\rangle
 - \langle \pi^0|T\{J^{\rm EM}_\mu,J^{\rm EM}_\nu\}|\pi^0\rangle \Big].
\label{eq:def_del_mpi2}
\end{eqnarray}
where $D_{\mu\nu}(q)$ is the photon propagator and
$J_\mu^{\rm EM}=\sum_f e_f\bar \psi_f\gamma_\mu\psi_f$ the EM current.
Using the soft-pion relation and the current algebra,
eq.(\ref{eq:def_del_mpi2}) can be written as
\begin{eqnarray}
  \Delta m_\pi^2 &=& \frac{e^2}{f_\pi^2}\int\frac{d^4 q}{(2\pi)^4} D_{\mu\nu}(q)\int d^4 x e^{iqx}
  \Big[ \langle 0|T\{V^3_\mu,V^3_\nu\}|0\rangle(x)
      - \langle 0|T\{A^3_\mu,A^3_\nu\}|0\rangle(x) \Big] \\
  &=& -\frac{3\alpha^{\rm EM}}{4\pi f_\pi^2}\int^{\infty}_0 dQ^2 Q^2 \Pi_{V-A}(Q^2),
\label{eq:def_del_mpi2-2}
\end{eqnarray}
where $Q^2=-q^2>0$, $V_\mu^a=\bar\psi\gamma_\mu T^a\psi$ and
$A_\mu^a=\bar\psi\gamma_\mu\gamma_5 T^a\psi$ with SU(2) generator $T^a$
normalized by ${\rm tr}T^aT^b=\delta^{ab}$ \cite{DGMLY,Harada}.
Hereafter $f_\pi=130.5$ MeV normalization is adopted.
The second equation can be derived by substituting
$\langle 0|T\{J_\mu J_\nu\}|0\rangle=(\delta_{\mu\nu}Q^2-Q_\mu Q_\nu)\Pi_J(Q^2)$ and 
$D_{\mu\nu}=(\delta_{\mu\nu}Q^2-(1-\xi)Q_\mu Q_\nu)/Q^4$ with arbitrary gauge parameter $\xi$
into the first equation, and we define $\Pi_{V-A}=\Pi_V-\Pi_A$.
It should be noted that, in deriving eq.~(\ref{eq:def_del_mpi2-2}),
the chiral limit is taken after applying the soft-pion relation.
Thus the above is the exact formula in the chiral limit.

CHPT at one-loop order predicts the low momentum behavior of $\Pi_{V-A}(Q^2)$ \cite{GL} as
\begin{equation}
  \Pi_{V-A}(Q^2) = -\frac{f_\pi^2}{Q^2+m_\pi^2} - 8L_{10}^r(\mu_\chi) +
  \frac{1}{24\pi^2}\Big[ -\frac{1}{3} +
  \sigma^2\Big(\sigma\ln\frac{\sigma-1}{\sigma+1}+2\Big)-\ln\frac{m_\pi^2}{\mu_\chi^2}\Big]
 + \mathcal O(Q^4),
\label{eq:CHPT_Pi_V-A}
\end{equation}
with $\sigma=\sqrt{1+4m_\pi^2/Q^2}$.
$L_{10}^r$ is related to the S-parameter \cite{Peskin}, which plays an
important role in analyzing new physics models.
On the other hand, at large momentum OPE \cite{OPE2} provides
\begin{equation}
   \Pi_{V-A}(Q^2) \simeq \frac{C_{d=2}m_q^2(Q^2)}{Q^2} 
   + \frac{C_{d=4}m_q\langle \bar \psi\psi\rangle}{Q^4} 
   + \frac{C_{d=6}}{Q^6} + \mathcal O(Q^{-8})
\label{eq:OPE_Pi_V-A}
\end{equation}
where the explicit form of $C_{d=2,4}$ has been known to two-loop order and
$C_{d=6}$ is given by
\begin{equation}
 C_{d=6} = 8\pi\langle \alpha_s O_8\rangle + (\textrm{log term}) + \mathcal O(\alpha_s^2),
\end{equation}
and
\begin{equation}
\langle O_8\rangle_{\mu_o} = \Big\langle 
    (\bar\psi\gamma_\mu \lambda^\alpha\frac{t^3}{2}\psi)(\bar\psi\gamma^\mu \lambda^\alpha\frac{t^3}{2}\psi)
  - (\bar\psi\gamma_\mu \gamma_5 \lambda^\alpha\frac{t^3}{2}\psi)
    (\bar\psi\gamma^\mu \gamma_5 \lambda^\alpha\frac{t^3}{2}\psi)
  \Big\rangle_{\mu_o}.
\label{eq:ope formula}
\end{equation}
$\lambda^\alpha$ is the Gell-Mann color matrix and $t^a$ is the Pauli matrix, 
and $\mu_o$ is a renormalization scale.
The logarithmic term is estimated to be a few \% of the leading term, which we ignore.
$\langle O_8\rangle$ contains information of the matrix element of
$K^0\rightarrow (\pi\pi)_{I=2}$ \cite{Donoghue}.

\subsection{Lattice formula}
In this simulation we use the following vector and axial-vector currents,
\begin{equation}
  V_\mu = Z_V\bar\psi\gamma_\mu\Big(1-\frac{D_{ov}}{2m_0}\Big)\psi,\quad
  A_\mu = Z_A\bar\psi\gamma_5\gamma_\mu\Big(1-\frac{D_{ov}}{2m_0}\Big)\psi
\label{eq:V_A}
\end{equation} 
where $m_0=1.6$ and non-perturbative renormalization constant
$Z_V=Z_A=1.38$ \cite{Noaki} is applied.
Since these currents are not the conserved one, the current-current
correlation functions may contain lattice artifacts.
Their explicit form is represented by
\begin{eqnarray}
   \langle 0|T\{J_\mu,J_\nu\}|0\rangle 
&=& \Big(\delta_{\mu\nu} Q^2 - Q_\mu Q_\nu\Big) \Pi^{(1)}_J(Q^2)
   - Q_\mu Q_\nu \Pi^{(0)}_J(Q^2) 
   + A(Q^2)\delta_{\mu\nu} + (aQ_\mu)^2\delta_{\mu\nu}B_1(Q^2) 
\nonumber\\
&+& (aQ_\mu)^4\delta_{\mu\nu} B_2(Q^2) + \Big((aQ_\mu)(aQ_\nu)^3 
+ (aQ_\mu)^3(aQ_\nu)\Big)C_{11}(Q^2) + \cdots
\label{eq:JJ_lat}
\end{eqnarray}
The ellipsis denotes the higher order terms and hereafter we ignore these terms.
In order to remove the lattice artifacts $A,B_1,B_2,C_{11}$ we impose the
Ward-Takahashi identity by constructing a set of 
linear equations, $Q_\mu\langle J_\mu J_\nu\rangle=f_1(\Pi_J,A,B,C)$, 
$Q_\mu Q_\nu\langle J_\mu J_\nu\rangle=f_2(\Pi_J,A,B,C)$, $\cdots$, 
for different momentum configurations giving the same $Q^2$.
By solving these linear equations, we obtain $\Pi_J=\Pi_J^{(0)}+\Pi_J^{(1)}$.

\section{Numerical results and analysis}
\subsection{Lattice parameters}
We use the $N_f=2$ dynamical overlap fermion configurations with the Iwasaki gauge action 
at $\beta=2.3$ corresponding to $a^{-1}=1.67$ GeV on a $16^3\times 32$ lattice 
generated at a fixed topological charge $Q^{\rm top}=0$ \cite{Hashimoto}. 
The sea quark masses are chosen to $m_q=0.015,\,0.025,\,0.035,\,0.050$ 
and the valence quark mass takes the same values as the sea.
The pion mass squared are then $0.082,\,0.134,\,0.189,\,0.27$ GeV$^2$.
The number of statistics is 200 configurations separated by 50 HMC trajectories.
The statistical error is estimated using the jackknife method with bin size equal to 2.

\subsection{Chiral extrapolation}
\begin{figure}
\begin{center}
\includegraphics[width=75mm]{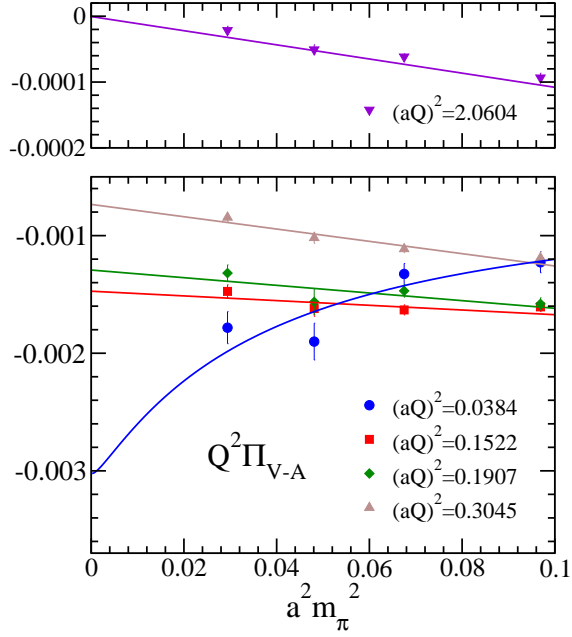}
\end{center}
\vskip -5mm
\caption{The dependence of pion mass squared for $Q^2\Pi_{V-A}$ at 4 low momenta, and 
1 high momentum. The straight lines denote the fit function. Except for the lowest momentum 
fit function is linear function.}
\label{fig:Q2Pi_V-A_m}
\end{figure}

The chiral extrapolation of $\Pi_{V-A}(Q^2)$ is made at each momenta.
The $(a^2 m_\pi^2)$ dependence and fit results are shown in
Fig.\ref{fig:Q2Pi_V-A_m} for several representative $(aQ)^2$'s.
At the lowest momentum, $(aQ)^2=0.0384$, the CHPT prediction
(\ref{eq:CHPT_Pi_V-A}) is used together with $f_\pi^2$ and $m_\pi^2$ obtained from the
pseudoscalar two-point correlation function.
As shown in Fig.\ref{fig:Q2Pi_V-A_m} the fit describes the data reasonably well, and leads to
\begin{equation}
  L_{10}^r(\mu_o=770\,{\rm MeV}) = -0.00474(23),
\end{equation}
which is in good agreement with experimental value 
$L_{10}^r(\mu_o=770\,{\rm MeV})|^{\rm Exp.}=-0.00509(47)$ \cite{Ecker2}. 
For other momenta, the data show linear behavior, that we fit with a linear function.
This implies that the quadratic quark mass term in
eq.(\ref{eq:OPE_Pi_V-A}) is small, and the dimension-four and dimension-six
terms are dominating in small quark mass region.

We also try to estimate the four-quark condensation $\langle O_8\rangle$
by fitting to a functional form similar to eq.~(\ref{eq:OPE_Pi_V-A}),
\begin{equation}
  \Pi_{V-A}(Q^2) = \frac{D_1m_q^2(Q^2)}{Q^2} + \frac{D_2m_q}{Q^4} + \frac{D_3m_q+D_4}{Q^6}
\end{equation}
where the fit parameters $D_{1}$, $D_{2}$  and $D_3m_q+D_4$ correspond to
$C_{d=2}$, $C_{d=4}\,\langle\bar\psi\psi\rangle$ and $C_{d=6}$,
respectively, however any $Q^2$ dependence in $C_{d=2,4,6}$ is omitted.
Fitting range is chosen as $[1.235,1.973]$ where OPE is expected to be
a dominant contribution to the vacuum polarization.
In the chiral limit, $D_4=\langle 8\pi\alpha_s O_8\rangle$ up to
a logarithmic correction, and we obtain 
\begin{equation}
  \langle O_8\rangle_{\mu_o=2\,{\rm GeV}} = -0.20(13) \times 10^{-3}\,{\rm GeV}^3
\label{eq:O_8}
\end{equation}
with $\alpha_s(\mu_o=2\,{\rm GeV})=0.334$, which is used in \cite{Donoghue}.
In Ref.\cite{Donoghue}, their estimated value in the $\overline{\rm MS}$ scheme is
$-(0.67\sim 1.29)\times 10^{-3}$ GeV$^{-3}$, which was the same sign with our result while 
whose magnitude is larger than eq.(\ref{eq:O_8}).
Although our result still contains a large uncertainty, 
it suggests a feasibility to extract the expectation values appearing in the
QCD sum rule analysis.

\subsection{Numerical integral}
\begin{figure}
\begin{center}
\includegraphics[width=100mm]{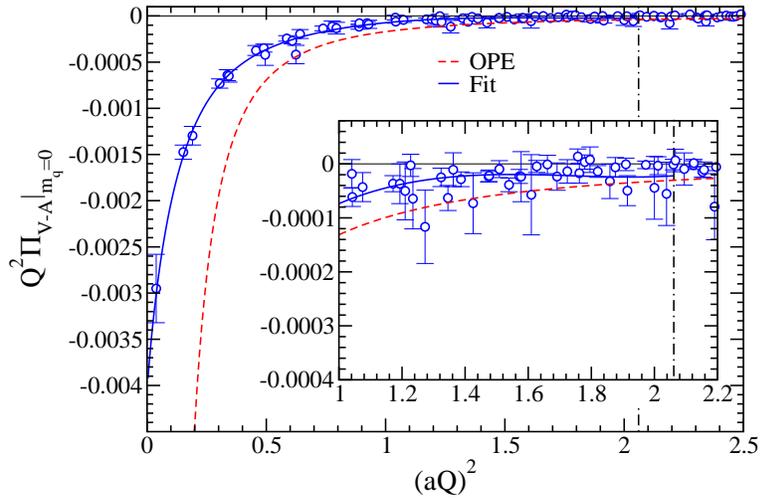}
\end{center}
\vskip -5mm
\caption{Momentum dependence of $Q^2\Pi_{V-A}(Q^2)$ in the chiral limit.
The dashed line denotes the prediction of OPE at leading order, and straight line denotes fit function.
The dashed-dots line is a cutoff point.}
\label{fig:Q2Pi_V-A_m0}
\end{figure}

After chiral extrapolation we obtain the momentum dependence of $Q^2\Pi_{V-A}$ in the chiral limit 
as shown in Fig.\ref{fig:Q2Pi_V-A_m0}. In order to perform the numerical integral for 
$Q^2\Pi_{V-A}$, we fit with an appropriate function. 
From eq.~(\ref{eq:CHPT_Pi_V-A}), $Q^2\Pi_{V-A}$ in the massless limit is given
by 
\begin{equation}
  \lim_{m_\pi^2\rightarrow 0}Q^2\Pi_{V-A}(Q^2) = 
  -f_\pi^2 -\frac{Q^2}{24\pi^2}\ln\frac{Q^2}{\mu_\chi^2} + \mathcal O(Q^2), 
\label{eq:chpt_mq0}
\end{equation}
which should be satisfied when making the fit ansatz.
Furthermore, $\Pi_{V-A}$ contains poles corresponding to resonance states at
negative $Q^2$. 
According to these requirements and the OPE prediction, we take the following
fit function
\begin{equation}
  Q^2\Pi^{\rm fit}_{V-A}(Q^2) = -F^2 + \frac{Q^2F_1^2}{Q^2+M_1^2} - \frac{Q^2F_2^2}{Q^2+M_2^2}
  + \Big(-\frac{Q^2\ln Q^2}{24\pi^2}+c_1 Q^2\Big)\frac{1}{1+c_2 Q^6}.
  \label{eq:fit function}
\end{equation}
$M_{1,2},\,F,\,F_{1,2},\,c_{1,2}$ are free parameters.
The last term expresses the logarithmic term at small $Q^2$ and this is
suppressed by $c_2Q^6$ for large $Q^2$.
Figure \ref{fig:Q2Pi_V-A_m0} shows that the fit function
(\ref{eq:fit function}) well describes our data below $(aQ)^2=2.0604$. 
From the fitting result of $F^2$,
\begin{equation}
  f_\pi = 107(15) \,{\rm MeV}
\label{eq:f_pi}
\end{equation}
is obtained.
This value is consistent with $f_\pi\sim 110$ MeV, which is
obtained from the study of hadron spectrum using the
same configurations \cite{Noaki}.

In the numerical integral we split the integral range into two regions at $Q^2=\Lambda^2$:
\begin{equation}
  \Delta m_\pi^2 = -\frac{3\alpha^{\rm EM}}{4\pi^2 f_\pi^2}\Big[ 
  \int^{\Lambda^2}_0 dQ^2 Q^2 \Pi^{\rm fit}_{V-A}(Q^2) 
+ \int^\infty_{\Lambda^2} dQ^2 Q^2 \Pi^{\rm OPE}_{V-A}(Q^2)
  \Big]
\label{eq:delta_mpi}
\end{equation}
Below $\Lambda^2$ we use the fit function (\ref{eq:fit function}) and
the parameters determined by the fit, while above $\Lambda^2$ we employ the perturbative form
at the one-loop order \cite{OPE2} with the factorization method for 
the expectation value of $O_8$, 
\begin{equation}
Q^2\Pi_{V-A}^{\rm OPE} = -\frac{64\pi}{9}\alpha_s(\mu_o)\langle\bar\psi\psi\rangle^2\Big[ 
  1 + \frac{\alpha_s(\mu_o)}{\pi}\Big(\frac{89}{48} - \frac{1}{4}\ln\frac{Q^2}{\mu_o^2}\Big)\Big]Q^{-4}
\end{equation}
with $\mu_o=2$ GeV and $\langle\bar \psi\psi\rangle=-(251\,{\rm MeV})^3$ \cite{Fukaya}.
$\Lambda^2$ is set to $a^2\Lambda^2=2.0604$, where the gap between the two regions 
is negligibly small. Our result is
\begin{equation}
  \Delta m_\pi^2 = 976(100)_{\rm stat.} + 48_{\rm OPE}\,{\rm ~MeV^2}
  = 1024(100)\,{\rm ~MeV^2},
\end{equation}
where we insert $f_\pi$ in eq.(\ref{eq:f_pi}) into the denominator of eq.~(\ref{eq:delta_mpi}).

In addition to the statistical error given above, there are several sources of
systematic error.
Since the physical volume of our lattice is about (1.9 fm)$^3$, the
lightest pion data ($\sim$290 MeV) could receive sizable finite size effect.
Also, since this result is obtained at a fixed topological sector
($Q^{\rm top}=0$), an additional finite size effect of $O(1/V)$ is expected
\cite{Aoki}.
We may include such effects by modifying the chiral extrapolation.
In the comparison to the experimental value, it is also necessary to evaluate
the correction due to the small but finite quark masses.
Although these systematic errors are yet to be estimated, it is encouraging
that our result is reasonably consistent with the experimental value,
$\Delta m_\pi^2({\rm Exp.})=1261.2$ MeV$^2$.

\section{Summary}
In this work we have calculated the pion mass difference by applying the
DGMLY sum rule to the lattice calculation for the first time.
We also obtained $f_\pi$ and $L_{10}^r$ by comparing the $Q^2$ dependence of the vacuum
polarization with the predictions of CHPT.
The use of the overlap fermion made this calculation possible.

\section{Acknowledgments}
We thank M. Golterman for useful discussions.
This work is supported by the Grant-in-Aid of the Japanese Ministry of Education
(No. 18034011, 
     18340075, 
     18740167, 
     18840045, 
     19540286, 
     19740121, 
     19740160  
).
Numerical simulations are performed on Hitachi SR11000 and IBM System Blue
Gene Solution at High Energy Accelerator Research Organization (KEK) under a
support of its Large Scale Simulation Program (No.~07-16).

\end{document}